\let\accentvec\vec
\documentclass[american]{llncs}

\let\vec\accentvec

\usepackage{underscore}           

\usepackage{babel}
\usepackage[final]{graphicx}

\usepackage[utf8]{inputenc}
\usepackage{amsmath}
\usepackage{mathtools}[2011/02/12]
\usepackage{amsfonts,amssymb,stmaryrd}
\usepackage[retainmissing]{MnSymbol}

\usepackage[final]{listings}
\usepackage{tikz}
\usetikzlibrary{positioning}

\usepackage[bookmarks=true,bookmarksopen=true,pdfhighlight=/I,pdfpagemode=UseOutlines,backref=false,final]{hyperref}
\usepackage[fancyexamples,fancyproofs,extended]{theorems}[2016/07/22]

\input{wgmacros-FLP}
\renewcommand*{\Psyms}{\mathit{PId}}

\newsmartprefixextended{eqs}{\equationsname}{(}{)}
\newsmartprefix{line}{Line}

\makeparensmathoper{\RunAtom}{\overleftarrow{\textbf{run}}}
\makeparensmathoper{\Run}{\textbf{run}}
\makeparensmathoper{\atomicsAux}{fixAt}
\makeparensmathoper{\atomics}{fixAtomics}
\makeparensmathoper{\interlvAux}{unroll'}
\makeparensmathoper{\interlv}{unroll}
\makeparensmathoper{\propagate}{prpgt}
\newcommand*{\CondStepAtom}[2]{ #1 \mathrel{\overleftarrow{\medtriangleright}} #2  }
\newcommand*{\CondStep}[2]{ #1 \medtriangleright #2  }
\newcommand*{\SemProc}[3][\Pi_{p}]{\syntaxoper{\mathcal{D}}{#1}{#3}^{#2}}
\newcommand*{\SemTP}[4][\Pi_{p}]{\syntaxoper{\mathcal{P}}{#1}{#3,#4}^{#2}}
\newcommand*{\arcElse}[3][l]{#1 \xRightarrow{#2} #3}
\newcommand*{\arcs}[1][p]{\mathcal{E}_{#1}}
\newcommand*{\arc}[3][l]{#1 \xrightarrow{#2} #3}
\newcommand*{\cond}[1]{#1^C}
\newcommand*{\labs}[1][p]{\mathcal{L}_{#1}}
\newcommand*{\marked}{\overleftarrow}
\newcommand*{\trans}[1]{#1^T}

\newcommand*{\CT}{\mathcal{CT}}

\newcommand*{\Clang}{\textsf{C}}

\newcommand*{\In}{\I[]}
\newcommand*{\LTL}{\textsf{LTL}}

\newcommand*{\SemLEval}[3]{\syntaxoper{\mathbb{L}}{#1}{#3}^{#2}}
\newcommand*{\SemProcs}[3]{\syntaxoper{\mathbb{D}}{#1}{#3}^{#2}}
\newcommand*{\SemProg}[2]{\syntaxoper{\mathbb{P}}{#1}{#2}}
\newcommand*{\SemREval}[3]{ \syntaxoper{\mathbb{R}}{#1}{#3}^{#2}} %

\newcommand*{\SemStmt}[3]{\syntaxoper{\mathbb{S}}{#1}{#3}^{#2}}

\newcommand*{\State}[0]{ \textit{State} }
\newcommand*{\Stop}{ \blacksquare }

\newcommand*{\predef}[1]{\ensuremath{\mathtt{\_#1}}}
\newcommand*{\promela}{\textsf{PROMELA}}
\newcommand*{\spin}{\textsf{SPIN}}

\newcommand*{\update}[2]{\big[ #2 \big/ #1 \big]}
\title{A denotational semantics for \promela{} addressing arbitrary jumps}
\author{Marco Comini\inst{1}\orcidID{0000-0002-8069-3411} \and
Mar\'\i a del Mar Gallardo\inst{2}\orcidID{0000-0003-3481-5307} \thanks{This author has been partially supported by the Spanish MCIU under grant
RTI2018-099777-B-I00.}\and
Alicia Villanueva\inst{3}\orcidID{0000-0003-1090-5009}\thanks{This author has been partially supported by the EU (FEDER) and the Spanish MCIU under grant
RTI2018-094403-B-C32, and by Generalitat Valenciana under grant PROMETEO/2019/098.}}
\authorrunning{M. Comini et al.}
\institute{Universit\`a degli Studi di Udine (Italy)\\
\email{marco.comini@uniud.it} \and 
Universidad de M\'alaga, Andaluc\'{\i}a Tech, \\
ITI-Software\\
\email{mdgallardo@uma.es} \and
VRAIN - Universitat Polit\`ecnica de Val\`encia\\
\email{alvilga1@upv.es}}

\begin{document}
\maketitle

\begin{abstract}
    \promela{} (Process Meta Language) is a high-level specification
    language designed for modeling interactions in distributed
    systems.  \promela{} is used as the input language for the model
    checker \spin{} (Simple Promela INterpreter).  The main
    characteristics of \promela{} are non-determinism, process
    communication through synchronous as well as asynchronous
    channels, and the possibility to dynamically create instances of
    processes.

    In this paper, we introduce a bottom-up, fixpoint semantics that
    aims to model the behavior of \promela{} programs.  This work is
    the first step towards a more ambitious goal where analysis and
    verification techniques based on abstract interpretation would be
    defined on top of such semantics.
\end{abstract}

\section{Introduction}

Concurrent programming has posed additional difficulties to the
problem of guarantying correctness of programs.  In fact, it may be
almost impossible to replicate and debug an error caught by testing
due to the fact that the different threads or processes can interact
with each other at a different pace (depending on, for instance,
processors speed).  Not to mention the lack of coverage that usually
testing techniques suffer.

Program verification is needed to ensure the correctness of a program
in the case when, first, testing is simply not reliable enough due to
a large number of possible inputs and, second, the consequences of
failure are too costly in terms either of economic losses or of
consequences that affect our lives.

In this context, \promela{} and \spin{} (\cite{Holzmann04}) provide
the tools for effectively model check
(\cite{Clarke-Emerson-1982,Queille-Sifakis-1982}) the concurrent
behavior of systems (specified in \promela{}).  \spin{} incorporates
powerful techniques to handle big (finite) search spaces, some of them
based on approximations or on heuristic searches.  However, no
complementary techniques (different from model checking) for the
formal verification of \promela{} models have been proposed in the
literature.  A tool to check some of the classical equivalences
defined for \promela{} is presented in \cite{Erdogmus02}.

This work constitutes the first step of a project that aims to provide
an approach to verification of properties of \promela{} programs
different from model checking.  More specifically, we are interested
in analysis and verification techniques based on approximation via
abstract-interpretation techniques.  Depending on the kind of
technique to be used, denotational semantics with some properties can
be more convenient than others.  On top of the concrete inductive,
fixpoint semantics defined in this work, it will be possible to define
appropriate abstract semantics which implement specific analysis, for
instance related to the temporal behavior of the system.
The experience with the semantics proposed for the \textsf{tccp}
language \cite{CominiTV13cltl} showed that good results are obtained
when abstraction is based on a concrete inductive, fixpoint semantics.
To the best of our knowledge, in the literature, we only find
proposals for (transitional) operational semantics for \promela{}.

The semantics defined in this paper works in two phases.  Intuitively,
at a process level we first collect for each process declaration a set
of hypothetical traces corresponding to the set of
behaviors \emph{independent from the context}.  As one can imagine,
some of these traces may become \emph{impossible to travel} given a
context (a set of processes run in parallel).
Then, in a second phase at a system level, the traces from the
processes running in parallel are combined to obtain \emph{global}
traces.  During this phase, some traces might be discarded (for
instance, when a synchronous communication is requested but there is
no process able for \emph{handshaking}).

Due to the \promela{} language spirit, our plan is to focus on global
(concurrent) properties, namely on defining analyses on top of the
sets obtained at the system level.  We note that it would be possible
to define analyses for local processes (on top of the sets obtained
after the first phase) for analyzing the behavior of local
variables. However, these would be in general of less interest since,
1) \promela{} defines concurrency only at the level of the system (in
contrast to other concurrency models, there is no parallel operator
among the basic constructs), and 2) the language is strongly designed
to precisely capture the concurrent behavior and abstract as much as
possible the local behavior of processes (as a means to reduce the
number of states to explore).

One of the purposes of the \promela{} language is to facilitate the
modeling of concurrent state machines.  This is why it includes
the \texttt{goto} statement that is intensively used by programmers to
implement {\em jumps} from source to target states.  It is well known
that dealing with arbitrary jumps constitutes a technical difficulty
when defining a semantics by induction on syntax.  In this paper, for
the sake of conciseness, we build the process denotations based on the
static graph provided by the process code.

This work is organized as follows.  In the following section, we
introduce the \promela{} language.  \smartref{sec:semanticsDomain} is
devoted to the presentation of the denotational domain.  The
formalization of the fixpoint denotational semantics and also some
examples are shown in \smartref{sec:denotational}.  Finally, we
conclude by discussing some relevant properties of the semantics and
describe our plans for future work.

\section{Promela}\label{sec:promela}

\promela{} is a high-level specification language designed for modeling interactions in
concurrent and distributed systems \cite{Holzmann04}.  Its
specification is given on the
\spin{} website\footnote{\url{http://spinroot.com}}.  Other important sources are
\cite{Holzmann12,Benari08}.
Given a system specified in \promela{}, \spin{} can either simulate
the system's behavior or it can generate a \Clang{} program that
carries out the verification of the system against some specified
correctness properties.  These properties can be defined within the
model or as temporal \LTL{} properties (\cite{Pnueli-1971}).
In order to achieve effectiveness, \promela{}'s core design philosophy
is to limit the size of the state space of models.  This is enforced
by introducing some restrictions on the language and trying to create
a model where the optimal model abstraction is desired (\eg{}, by
avoiding both to underspecify and to overspecify the system).
However, to consider only finite sized models is often not enough to
overcome the state-explosion problem.

The emphasis of the language is not on computation details but on the
modeling of process synchronization and coordination.
Therefore, some common features of implementation languages such as
input-output constructs or floating-point arithmetic are just not
present in \promela{}.  In contrast, \promela{} is non-deterministic,
allows for the dynamic creation of concurrent processes and processes
can communicate via message channels that can be either synchronous or
asynchronous using the CSP notation \cite{Hoare85}.

\subsection{Syntax and informal operational semantics of \promela{}}\label{sec:syntax}
For space reasons, we do not describe the syntax and operational
semantics of \promela{} constructors exhaustively in this paper.
Instead, we give an intuitive description of the main features of
the \promela{} language by means of examples.  As it will be clear in
the following discussion, the syntax of the language is similar
to \Clang{}.
A \promela{} program consists of a sequence of global declarations
(types and variables), process declarations (\texttt{proctype}) and
the (optional) \texttt{init} process used to initialize the system,
when required.

\begin{figure}[tp]
{\small
    \begin{lstlisting}
        bit f[2] = {0,0};|\label{line:dec}|
        proctype P(bit id){|\label{line:proc}|
          L0: if|\label{line:L0}| :: atomic{ f[id] = 1;|\label{line:atomic}|
          L1:        !f[1-id] } ;|\label{line:L1}|
          L2:       skip; //critical section
          L3:       f[id] = 0; 
          L4:       goto L0|\label{line:L3}|
              fi|\label{line:endif}| }|\label{line:endproc}|
        init{|\label{line:init}| L5: atomic{ run P(0); L6: run P(1) }|\label{line:runs}| }|\label{line:endinit}|
    \end{lstlisting}
    }
    \vspace{-1ex}
    \caption{Example of \promela{} code}\label{fig:peterson}
\end{figure}
Consider as running example the \promela{} code given
in \smartref{fig:peterson}.  The program corresponds to a variant of
the well-known \emph{first attempt} to the Peterson's
algorithm \cite{Peterson81} to solve the mutual exclusion problem for
two processes.  The algorithm contains two processes \texttt{P(0)}
and \texttt{P(1)} which are trying continuously to enter in its
critical section (CS).  To avoid to violate the mutual exclusion
property (that is, to ensure that always there is at most one process
executing its critical section), each process uses a boolean flag
(\texttt{f[0]} and \texttt{f[1]}).
When a process wants to enter its CS, it changes its own boolean to
true and waits for the other process to be outside its CS (its boolean
is false).  After executing the CS, each process sets its own boolean
to false.  Although this first attempt solves the mutual exclusion
requirement, it contains many possible deadlocked executions.

As can be observed, the program starts by defining,
in \smartref{line:dec}, a global array
\texttt{f} with two components of the predefined type \texttt{bit}.  Both components are
initialized to false (\texttt{0}).  Next, in
Lines~\ref{line:proc}-\ref{line:endproc}, the
\texttt{proctype P} is declared.  \texttt{Proctypes} are similar to \Clang{} functions (they
can contain local types, variables and executable code) except for,
when invoked, their code is always executed in a new execution thread.
When processes do not need initial values,
\promela{} allows to directly declare and instantiate them placing the keyword \texttt{active}
before \texttt{proctype}.

The \texttt{init} process in Line~\ref{line:init} acts as
the \texttt{main} function in
\Clang{}.  In our example it is used to create by means of statement \texttt{run} two instances
of the \texttt{proctype P}, one with \texttt{id=0} and the other
with \texttt{id=1}.  When processes are created, \spin{} identifies
them through its \emph{pid} which is an integer different for each
process instance.

The evolution of a \promela{} program proceeds by freely interleaving
the instructions of the processes in execution.  To completely
understand this behavior is mandatory to introduce the notion
of \emph{executability} in \promela{}.  \spin{} selects an instruction
to be next executed only if the instruction is an executable
instruction of a process, that is, a statement that does not suspend.
If no executable statement can be selected, the system blocks.
Assignments (as \texttt{f[id]=1} in \smartref{line:atomic})
and \texttt{run} statements are always executable.  However, boolean
expressions (as \texttt{!f[1-id]} in
\smartref{line:L1}) are only executable when they are true.  This is the natural
synchronization mechanism based in shared memory implemented
in \promela{}.  A similar behavior occurs for instance when processes
communicate using shared channels.  The send or receive channel
operations are only executable if the corresponding channels are not
full or not empty, respectively.

The \texttt{atomic} statement in \smartref{line:runs} is used to avoid
that the \texttt{init} process loses the execution control during the
creation of the two new processes.  When a statement inside
an \texttt{atomic} statement is non-executable (as the boolean
expression of
\smartref{line:L1} when it evaluates to false), \spin{} breaks the atomic process execution to
avoid the system deadlock.  Thus, the \texttt{atomic} statement should
be read as ``execute the code in an atomic manner while it is
possible''.  An atomic statement is executable iff its first nested
statement is.  For instance,
\texttt{atomic} in \smartref{line:L0} is always executable since its first nested statement
\texttt{f[id] = 1} is an assignment that can be always be executed.

Channels are declared through variable declarations by using the
\texttt{chan} type.  The declaration \texttt{chan c = [N] of $\{$bit,mtype$\}$} defines a
channel {\tt c} of capacity {\tt N} for storing \texttt{N} records
composed of a \texttt{bit} and an \texttt{mtype} (enumerated types
that users can freely define).  The send and receive operations on
channels are written using the CSP notation.  For
instance, \texttt{c?x,y} and
\texttt{c!1,m} are operations on channel \texttt{c} defined above, where we are assuming that
\texttt{m} is a value of type \texttt{mtype}, and \texttt{x} and \texttt{y} variables of
appropriate type.
If the capacity \texttt{N} is greater than zero, a buffered channel is
 created.  Otherwise, a rendezvous channel (synchronous channel) is
 created.  Rendezvous channels can pass messages only through
 handshake communication between sender and receiver and cannot store
 messages.

The control statements in \promela{} are \texttt{do/od}
and \texttt{if/fi}, although
\texttt{do/od} is not really needed since it can be simulated with \texttt{if/fi} and
\texttt{goto}.  These statements introduce the \emph{non-determinism} in the language.  An
\texttt{if/fi} statement such as the one in Lines~\ref{line:L0}-\ref{line:endif} of
\smartref{fig:peterson} starts with the keyword \texttt{if}, ends with \texttt{fi} and contains
one or more branches that begin with symbols \texttt{::} (the
instruction in the example has only one branch).  The execution of
an \texttt{if} statement consists in the execution of one of its
branches.  To do this, \spin{} selects one of the executable branches
non-deterministically.  A branch is executable if its first
instruction (the guard) is executable.  If no branch is executable,
then the \texttt{if/fi} instruction suspends.  In the case of
the \texttt{if/fi} statement of \smartref{fig:peterson}, as its guard
(the
\texttt{atomic} statement of \smartref{line:L1}) is always executable, the selection never
suspends.  It is also possible to add an unique ending \texttt{else}
branch to statements
\texttt{if/fi}.  In this case, the \texttt{if/fi} never suspends, since when no branch is
executable the \texttt{else} branch is selected.

As we observe in \smartref{fig:peterson}, \promela{} statements can be
labeled using identifiers.  For instance, \texttt{L3} is the label of
instruction \texttt{f[id]=0}.

\paragraph{Operational engine.}\label{sec:semanticengine}

The official semantics of PROMELA is defined in terms of an
operational model which contains one or more processes, zero or more
variables, zero or more channels, and a \emph{semantics engine} to
handle the non-determinism (due to processes interleaving or to
selection statements) and some internal variables such
as \emph{exclusive} (for \texttt{atomic}) or
\emph{handshake} (for \emph{rendezvous}).
Other read-only predefined variables handled by the semantic engine
are \verb!_pid!, which is local to each process and contains the
corresponding process identifier, and \verb!_nr_pr!  which stores the
number of active processes in the system.

The local control of each process is represented by a labeled
transition system (LTS).  Each LTS models the local behavior of a
single process.
The semantics engine runs the system in a stepwise manner: selecting
and executing one basic transition from some LTS at the time.  In
brief, the behavior of the engine is as follows.  While there exist
executable (enabled) statements, it selects one of them to execute and
updates the global state.
There are some statements that might globally influence executability
and that are handled by means of other internal variables.  For
instance, a side effect of the execution of any statement in an atomic
block (except the last one) is to set a variable \emph{exclusive} to
the $\mathit{pid}$ of the executing process, thus preserving the
exclusive privilege to execute.  We leave the rest of \emph{non-local}
constructs (such as \emph{timeout} or \emph{unless}) as future work.

In the model, there is no explicit characterization of time.  The
semantics engine keeps executing transitions until no executable one
remains.  This happens if either the number of active processes is
zero, or when a deadlock occurs.  The interested reader can
consult \cite{Holzmann12} for the pseudo-code of this process.

\section{The semantic domain}\label{sec:semanticsDomain}

The definition of our denotational semantics is inspired by the
semantics for the \textsf{tccp} language of \cite{CominiTV13sem}.  In
this work, denotations are based on so-called hypothetical
computations; \ie\ traces of conditional steps that ideally model
actual computations that would be obtained by feeding them with
initial system states that validate every condition.

The semantics domain for \promela{} essentially is formed by sequences
of possible evolutions of the stored information (\ie\ the system
memory and the channels instances).  As such, we first need to
formalize how the information stored at each execution point is
represented (\smartref{sec:FDefinitions}) and then we formalize the
notion of traces of conditional evolutions (\smartref{sec:condTraces})
that are the basic constituent of our denotations.

\subsection{The system state}\label{sec:FDefinitions}

The \emph{system state} of a \promela{} program contains a
representation of the memory and channel instances. Due to space
limitations, we do not formally introduce the different components of
the system (basic types, arrays, records and their associated
operators) but we informally introduce on demand the needed notation.
In the following, we use $\mathbb{V}$ for elements that can be stored
in memory and $\mathit{Channels}$ for the denotations of channels.
\begin{definition}[System State]
    A system state of a PROMELA program is an element of the domain $\mathit{State} =
    \left([\mathit{Loc} \rightarrow \mathbb{V}_\bot] \times [\mathit{ChanId} \rightarrow
    \mathit{Channels}_\bot] \right)_\bot$ where $\mathit{Loc}$ are locations of a memory
    capable to store denotable values $\mathbb{V}$, that are basic values, structures, and
    arrays\footnote{For all set $\mathbb{D}$, we write $\mathbb{D}_\bot$ for the flat domain of
    values of $\mathbb{D}$ with bottom $\bot$.}, and $\mathit{ChanId}$ is the set of channel
    ids; $\mathit{Channels}$ is the set of all channel instances;
    $\bot$ denotes an erroneous (inconsistent) state.
\end{definition}
In the sequel, we write $\sigma=(\sigma_L,\sigma_C)$ as a typical
element of $\mathit{State}$.  Moreover, for $l\in\mathit{Loc}$ we
write $\sigma(l)$ as a shorthand of $\sigma_L(l)$ or $\bot$ when
$\sigma=\bot$.  Analogously for $c\in\mathit{ChanId}$.

As usual, the scope is managed by environments, which are not required to be part of the state
since \promela{} has static scope.
\begin{definition}[Environment]
    An environment is a partial function associating program
    identifiers $\mathit{Ident}$ with locations or channel ids
    $\mathit{Env} =
    [\mathit{Ident} \rightharpoonup \mathit{Loc} \cup \mathit{ChanId}]$.
\end{definition}
In the sequel, we write $\rho$ as a typical element of the set of
$\mathit{Env}$.

\subsection{The Domain of Denotations}\label{sec:condTraces}

Let us now define the domain upon which our denotations are built.
Intuitively, each component in the semantics, called \emph{conditional
trace} aims to represent a sequence of hypothetical (conditional)
computational steps.

\begin{definition}[Conditional step and conditional traces]
    A \emph{conditional step} is either
    \begin{description} 
        
        \item[a conditional transition
        $\CondStep{p}{t}$]\vskip-\lastskip which consists of a state
        transformer $t : \State \rightarrow \State$ guarded by a
        condition $p : \State \rightarrow \mathit{Bool}$, or
             
        \item[a finalizer,] written $\Stop$, or
        
        \item[a process spawn,] written $\Run{\Psi}$, where $\Psi$ is
        a non-empty set of conditional traces.  \end{description}
    
    For conditional transitions and process spawns inside atomic
    blocks, we can add the markers $\CondStepAtom{p}{t}$ and
    $\RunAtom{\Psi}$.  In the sequel, we refer to the marked or
    non-marked version of conditional steps and process spawns
    indifferently, except in those cases when it is necessary to
    distinguish them.
    Given a conditional transition $\phi = \CondStep{p}{t}$ (or $\phi
    =\CondStepAtom{p}{t}$) we use the notation $\cond{\phi}$ and
    $\trans{\phi}$ to denote $p$ and $t$.

    A \emph{conditional trace} is defined as
    \begin{itemize}
        
        \item\vskip-\lastskip a finite sequence of conditional steps
        (except $\Stop$), possibly ending with $\Stop$,
        
        \item an infinite sequence of conditional steps (except $\Stop$).
    \end{itemize}

    We use $\varepsilon$ to denote the empty trace (of length zero).
    We denote by $\CT$ the set of all conditional traces.

    We denote by $\phi_1\cdot \ldots \cdot \phi_n$ the conditional
    trace whose conditional steps are $\seq{\phi}$.  Moreover, given a
    finite trace $\psi$ and a trace $\psi'$, we denote their
    concatenation by $\psi \cdot \psi'$.  Finally, given a set of
    traces $\Psi$, we denote by $\psi \cdot \Psi$ the set
    $\set{\psi \cdot \psi'}{\psi' \in \Psi}$.
\end{definition}
\begin{example}[Conditional trace]\label{ex:traces}
    Given an environment $\rho$, a trace formed by two conditional
    steps is $ \psi = \big((\CondStep{\lambda\sigma.  \True}
    {\lambda\sigma.  \sigma\update{\rho(\mathtt{x})}{\sigma(\rho(\mathtt{y}))}
    })\cdot \Stop \big)$.
The first step is a conditional transition whose predicate is always
    satisfied, and transformer binds variable $x$ to the value of
    variable $y$ in $\sigma$.  The second component of the trace is
    the finalizer $\Stop$.  Another example of conditional trace is
    $ \Run{\{\psi,\Stop\}}$.
\end{example}

Intuitively, a conditional transition $\CondStep{p}{t}$ represents all
possible computations where the current process state $\sigma$
satisfies the condition $p$ and then the process progress to state
$t(\sigma)$.
The finalizer $\Stop$ represents the end of the execution.
In the following section we show how the \emph{process spawn} is used
to handle process calls.

\begin{definition}[Preorder on $\CT$] 
    We order conditional traces by their information content.  Namely,
    for all conditional steps $\phi_1$ and $\phi_2$, for all
    $\phi_1 \cdot \psi_1$, $\phi_2 \cdot \psi_2$, $\psi\in\CT$ and for
    all $\Psi_1$, $\Psi_2 \subseteq \CT$, we define
    $\varepsilon \leq \psi$, $\Stop \leq \Stop$, and
\begin{align*}
        \phi_1 \cdot \psi_1 \leq \phi_2 \cdot \psi_2 &\iff \psi_1 \leq \psi_2 \wedge
        \forall\sigma \in \State .\, \cond{\phi_1}(\sigma) \implies \cond{\phi_2}(\sigma)
        \wedge \trans{\phi_1}(\sigma) = \trans{\phi_2}(\sigma) \\
        \Run{\Psi_1} \leq \Run{\Psi_2} &\iff \Psi_1 \sqsubseteq \Psi_2 \\
        & \text{ where } \Psi_1 \sqsubseteq \Psi_2 \iff \forall \psi_1 \in \Psi_1 \, \exists
        \psi_2 \in \Psi_2 .\, \psi_1 \leq \psi_2
    \end{align*}
\end{definition}
\begin{example}
    Given the conditional trace $\psi$ in \smartref{ex:traces} and the
    conditional trace $\psi'
    = \big(\CondStep{\lambda\sigma.\sigma(\rho(x))>0} {\lambda\sigma
    .\, \sigma\update{\rho(\mathtt{x})}{\sigma(\rho(\mathtt{y}))}
    }\big)\cdot \Stop$ we can observe that $\psi' \leq \psi$ since the
    predicate at the first position of $\psi'$ implies the one at the
    first position of $\psi$, and the remaining components are equal.
    This also means that the behaviors represented by $\psi'$ are also
    represented by $\psi$.
\end{example}

Note that $\sqsubseteq$ is not anti-symmetric; for instance, $\{ \phi \cdot \psi \} \sqsubseteq
\{ \phi, \phi \cdot \psi \}$ and $\{ \phi, \phi \cdot \psi \} \sqsubseteq \{ \phi \cdot \psi
\}$.  Therefore, in order to obtain a partial order, we use equivalence classes \wrt\ the
equivalence relation induced by $\sqsubseteq$.

\begin{definition}[Semantic domain]
    Given $M_1$, $M_2 \subseteq \CT$, we define $M_1 \Bumpeq M_2$ as
    $M_1 \sqsubseteq M_2 \wedge M_1 \sqsupseteq M_2$.
    We denote by $\C$ the class of non-empty sets of conditional
    traces modulo $\Bumpeq$.  Formally, $\C
    = \set{ \eqClass{M}{\Bumpeq} }{ M \subseteq \CT }$.
We abuse notation and denote by $\sqsubseteq$ also the partial order
    induced by the preorder $\sqsubseteq$ on equivalence classes of
    $\C$.

    $\Clattice$ is a complete lattice, where $\Ctop = \eqClass{\CT}{\Bumpeq}$, $\Cbot =
    \eqClass{ \{\varepsilon\} }{\Bumpeq}$ and, for all $\mathcal{M} \subseteq \C$, $ \Club{
    \mathcal{M} }{} = \eqClass{ \bigcup_{\eqClass{M}{\Bumpeq} \in \mathcal{M}} M }{\Bumpeq}$
    and $ \Cglb{ \mathcal{M} }{} = \eqClass{ \set{ \psi \in \CT }{ \forall \eqClass{M}{\Bumpeq}
    \in \mathcal{M} \, \exists \psi' \in M .\, \psi \leq \psi' } }{\Bumpeq}$
\end{definition}
It is worthy noting that we exclude $\emptyset$ from $\C$ since the
absence of progression is already represented by $\{\varepsilon\}$.

In the sequel, any $M \in \C$ is implicitly considered to be a set in $\CT$
that is obtained by choosing an arbitrary representative of the elements of
the equivalence class $M$.  Actually, all the operators that we use on $\C$
are independent of the choice of the representative.  Therefore, we can
define any operator on $\C$ in terms of its counterpart defined on sets of
$\CT$.

To give meaning to process calls we use the following notion of
interpretation that associates to each process symbol an element of $\C$
modulo variance.
\begin{definition}[Interpretations]
    \label{def:interp}
    Let $\Psyms$ be the set of process names and
    $\MGC[\Psyms] \dfn \{ \mgc{p}{l}{} \mid p\in \Psyms$, $\zseq{l}$
    are distinct locations$\}$ (or simply $\MGC$ when clear from the
    context).

    Two functions $I,J \colon \MGC[\Psyms] \to \C$
    are \emph{variants}, denoted by $I \FUNeq J$, if for each
    $\kappa \in \MGC[\Psyms]$ there exists a location remapping $\rho$
    such that $(I(\kappa))\rho = J(\kappa\rho)$.

    An \emph{interpretation} is a function
    $\I[] \colon \MGC[\Psyms] \to \C$ modulo variance\footnote{In
    other words, a family of elements of $\C$ indexed by $\MGC$ modulo
    variance.}.

    The semantic domain $\Cinterps^{\Psyms}$ (or simply $\Cinterps$ when clear from the
    context) is the set of all interpretations ordered by the point-wise extension of $\Cleq$
    (which by abuse of notation we also denote by $\Cleq$).
\end{definition}
Essentially, the semantics of each process in $\Psyms$ is given over
formal parameters that will be stored in a location that is actually
irrelevant.
It is important to note that $\MGC[\Psyms]$ has the same cardinality
of $\Psyms$ (and is thus finite) and therefore each interpretation is
a finite collection (of possibly infinite elements).

The partial order on $\Cinterps$ formalizes the evolution of the
process computation.  $\cpo{\Cinterps}{\CIleq}$ is a complete lattice
with bottom element $\lambda \kappa.\, \Cbot$, top element
$\lambda \kappa.\, \Ctop$, and the least upper bound and greatest
lower bound are the point-wise extension of $\Club{}{}$ and
$\Cglb{}{}$, \resp.  We abuse notation and use that of $\C$ for
$\Cinterps$ as well.

Any $\I[] \in \Cinterps$ is implicitly considered to be a function
$\MGC \to \C$ that is obtained by choosing an arbitrary representative
of the elements of $\I[]$ generated by $\FUNeq$.  Therefore, we can
define any operator on $\Cinterps$ in terms of its counterpart defined
on functions $\MGC \to \C$.
Moreover, we also implicitly assume that the application of an
interpretation $\I[]$ to a process call $\kappa$, denoted by
$\I[](\kappa)$, is the application $I(\kappa)$ of any representative
$I$ of $\I[]$ that is defined exactly on $\kappa$.

\section{The inductive bottom-up fixpoint semantics}\label{sec:denotational}

In the literature, there exist some works that introduce formal
definitions of an operational semantics for \promela{}.  For
instance, \cite{Weise97} proposes a clear structural operational
semantics defined in an incremental style based
upon \cite{NatarajanH96}.  However, such semantics models an old
version of the language and more recent features such as the new
behavior of atomic statement is missing.
In our opinion, the most relevant proposal is \cite{GallardoMP04},
which provides a parametric structural operational semantics defined
as the basis of the $\alpha$-\spin{} tool for abstract model checking.
In order to design our denotational semantics, we have took into
consideration only the official specification of \cite{Holzmann04}.
When any aspect was not defined there, we made our own assumptions and
then we validated them with the semantics in
\cite{Weise97,GallardoMP04} and with the latest version of \spin{}.

As the reader might have noted, we have restricted ourselves to a
significant fragment of
\promela{} and made some assumptions that we recapitulate here for clarity.
For simplicity, some constructs are not considered, for
instance \emph{timeout} or \emph{unless}, and the ordered insertion
and random removal of channel messages are not handled.
Moreover, we treat the primitive expression $\mathtt{run}$ as a
statement (and therefore it cannot be used into
expressions). Statements that can be translated into the included ones
(\mlst{od/do} or the \emph{for} statement) are also omitted.

The semantics is defined as the least fixpoint of a semantics operator $\SemProcs{}{}{} :
\Cinterps \to \Cinterps$ that transforms interpretations.  Informally, such fixpoint is
computed as the limit $\Club{ \set{
{\SemProcs{}{}{}}^{k}(\lambda \kappa.\, \Cbot) }{k\in\mathbb{N}} }{}$
where $\kappa \in \MGC[\Psyms]$.  Function $\SemProcs{}{}{}$ is
defined in terms of auxiliary semantics evaluation functions described
in the following sections.

\subsection{Denotation of Expressions} \label{sec:exprs}

Due to space limitations, we assume to have two functions to evaluate
expressions.  The first one is used for expressions on the right hand
side of an assignment and it returns the value of a given r-expression
in a given state $\SemREval{}{}{} : \mathit{r\mathord{-}expressions}
\times \mathit{Env} \times \mathit{State} \to \mathbb{V}$.  The second function, for
l-expressions, returns the modifier for the location associated to the
expression.  Given an environment $\rho$, a state $\sigma$ and then a
value-updating function $f:\mathbb{V} \to
\mathbb{V}$, the function $\SemLEval{}{}{} : \mathit{l\mathord{-}expressions} \times
\mathit{Env} \times \mathit{State} \to ((\mathbb{V} \to \mathbb{V}) \to \mathit{State})$ builds
a function to update the location identified by the l-expression with
modifier $f$.
\begin{example}[Evaluation of expressions]
    Given $\rho := \{ \mathtt{x} \rightharpoonup l_x
    \}$ and $\sigma := (\{ l_x \rightharpoonup 4 
    \}, \{\})$ , then
    $\SemREval{\mathtt{x + 1 > 3}}{\rho}{\sigma} = \SemREval{\mathtt{x +
    1}}{\rho}{\sigma} > \SemREval{\mathtt{3}}{\rho}{\sigma} =
    \SemREval{\mathtt{x}}{\rho}{\sigma} +
    \SemREval{\mathtt{1}}{\rho}{\sigma} > 3 = \sigma(\rho(\mathtt{x})) + 1
    > 3 = 1$ (which corresponds to $\True$).
\end{example}

\subsection{Denotation of Basic Statements of \promela} \label{sec:stmts}

We start by showing the denotation for basic statements.  Given a
basic statement $s$, an environment $\rho$ and an interpretation
$\In$, the function $\SemStmt{}{}{} : \mathit{stmt}
\times \mathit{Env} \times \Cinterps \to \CT$, written $\SemStmt{s}{\rho}{\In}$, builds the
conditional trace corresponding to the effects of $s$.
We show below the rules corresponding to the case when the constructs
are not within the scope of an atomic block. We use the variants
$\CondStepAtom{}{}$ and $\RunAtom{}$ replacing constructs
$\CondStep{}{}$ and $\Run{}$ in the given rules when the basic
statement is within the scope of an atomic block. The information
derived from the use of these variants will be used during the second
phase of the semantics computation (when synchronization among
processes is dealt with).
More specifically, following the operational behavior, the first
statement in an atomic block behaves as a guard to enter the atomic
block itself; if such guard is satisfied all statements of the block
will be run \emph{as an indivisible sequence} as far as they do not
suspend. There are some scenarios in which the statements inside an
atomic block may suspend however. In particular, when a synchronous
send statement occurs in an atomic block, the control is lost, thus
the behavior is equivalent to having an independent atomic block
starting at the statement right after the send statement.
Hence, as has been advanced, to later reproduce this behavior, when a
statement $s$ is inside an atomic block (except if it is the first one
of the block or it is the first one after a synchronous send
statement) the used operator is the one labeled with the arrow.

\begin{subequations}\label{eqs:SemStmt}
{\small
    \begin{align}
        & \SemStmt{\mathtt{goto}~L}{\rho}{\In} \dfn \varepsilon \label{eq:s:goto}\\
        & \SemStmt{\mathit{Expr}}{\rho}{\In} \dfn
        \CondStep{\lambda\sigma.\SemREval{\mathit{Expr}}{\rho}{\sigma}} {\id} \qquad\quad  \text{where $\id{}$ is the identity function.}
        \label{eq:s:condition} \\
        & \SemStmt{l = r}{\rho}{\In} \dfn \CondStep{\lambda\sigma.  \True}{\lambda\sigma.\,
        \SemLEval{l}{\rho}{\sigma}( \lambda v .  \SemREval{r}{\rho}{\sigma} )}
        \label{eq:s:assign}\\
        & \SemStmt{\mathit{type}\ x = e}{\rho}{\In} \dfn \CondStep{\lambda\sigma .\, \True}
        {\lambda\sigma.  \sigma \update{\rho(x)}{\SemREval{e}{\rho}{\sigma}} }
        \label{eq:s:basicDecInit} \\
        \shortintertext{For asynchronous channels $c$ (\ie\ whose capacity is
        greater than 0)}
        & \SemStmt{\mathit{ c!  \zseq{e}}}{\rho}{\In} \dfn
        \begin{aligned}[t]
            & \CondStep{\lambda \sigma .  \big(\sigma (\rho(c)) \neq \bot \wedge
            \mathoper{nfull}(\sigma(\rho(c)))\big) }{ \\*& \lambda\sigma .  
            \sigma\update{\rho(c)}{\mathit{push}( \sigma (\rho (c)),
            \SemREval{\zseq{e}}{\rho}{\sigma} )} }
        \end{aligned}
        \label{eq:s:sendAsync} \\
        & \SemStmt{\mathit{c?  \seq{x}}}{\rho}{\In} \dfn \CondStep{\lambda \sigma . \big( \sigma
        (\rho(c)) \neq \bot \wedge \mathoper{nempty}(\sigma(\rho(c)))\big) }{
        \label{eq:s:receiveAsync} \\*&\qquad\notag \lambda \sigma .  (
        \sigma\update{\rho(x_1)}{v_1} \dots \update{\rho(x_n)}{v_n}
        \update{\rho(c)}{\mathit{pop}(\sigma(\rho(c)))} )} \\*&\quad\notag\text{where }
        (\seq{v}) = \mathit{head}(\sigma(\rho(c))) \\
        \shortintertext{For synchronous channels $c$ (\ie\ whose capacity is 0)}
        & \SemStmt{\mathit{c!  \zseq{e}}}{\rho}{\In} \dfn
        \begin{aligned}[t]
            & \CondStep{\lambda \sigma .  \sigma (\rho(c)) \neq \bot }{ \\*& \lambda \sigma .
            (\sigma \update{\rho(c)}{\mathit{push}( \sigma (\rho (c)),
            \SemREval{\zseq{e}}{\rho}{\sigma} )}\update{\rho(\mathit{handshake})}{\rho(c)}) }
        \end{aligned}
        \label{eq:s:sendSync} \\
        & \SemStmt{\mathit{c?  \seq{x}}}{\rho}{\In} \dfn
        \CondStep{\lambda \sigma .  \big(\sigma (\rho(c)) \neq \bot \wedge
        \sigma(\rho(\mathit{handshake}))= \rho(c)\big) }{ \label{eq:s:receiveSync}
        \\*&\qquad\notag \lambda \sigma . 
        ( \sigma\update{\rho(x_1)}{v_1} \dots
        \update{\rho(x_n)}{v_n} \update{\rho(c)}{\mathit{pop}(\sigma(\rho(c)))}
        \update{\rho(\mathit{handshake})}{-1}) }
        \\*&\quad\notag\text{where } (\seq{v}) = \mathit{head}(\sigma(\rho(c)))
        \\
        %
        & \SemStmt{ \mlst{run} \ p\mlst{(} \zseq{e} \mlst{)}}{\rho}{\In} \dfn
        \big(\CondStep{\lambda\sigma.  \True}{\lambda \sigma .  (\sigma
        \update{\zseq{l}}{\zseq{v}} \update{l_{\mathit{pid}}}{\mathit{pid}}
        \update{\rho(\mathit{\predef{nr\_pr}})}{\sigma(\rho(\mathit{\predef{nr\_pr}})) +1})
        }\big)  \notag\\*&\qquad \cdot{} \Run{\In(p(\zseq{l}, \zseq{l_{\mathit{loc}}},
        l_{\mathit{pid}} )) } \label{eq:s:run} \\*&\quad\notag \noindent\text{where $\zseq{v} =
        \SemREval{\zseq{e}}{\rho}{\sigma}$, $\mathit{pid}$ is a fresh pid in $\sigma$ and
        $\zseq{l}, \zseq{l_{\mathit{loc}}}, l_{\mathit{pid}}$ are fresh locations in $\sigma$.}
    \end{align}
    }%
\end{subequations}%

\smartref{eq:s:condition} models the evaluation of an expression, whose right-evaluation is
used as the condition for the continuation of the trace.

\smartref{eq:s:assign} models an assignment (always enabled) and might alter the state
$\sigma$.

\smartref{eq:s:basicDecInit} model basic-type
variable initialization.  This action is always enabled and modifies
the system state.

\smartrefs{eq:s:sendAsync,eq:s:receiveAsync} model asynchronous communication.  The conditional
transition is executable only if the channel is currently defined and
not full (for sending) or not empty (for receiving).  The channel
valuation is updated as expected.  Functions
\emph{push}, \emph{pop} and \emph{head} are part of the semantic domain for system states.

\smartrefs{eq:s:sendSync,eq:s:receiveSync} model synchronous communication.  We treat
synchronous channels as channels with size one and synchronization is
achieved by using the predefined variable $\mathit{handshake}$.
The behavior for synchronous communication is similar to the
asynchronous one, the particular point is that the \emph{send}
statement updates the $\mathit{handshake}$ variable with the channel
identifier so that the \emph{receive} statement can check if the
channel it is listening to coincides with that in
$\mathit{handshake}$.  Upon completion, $\mathit{handshake}$ is set to
its default value ($-1$).  Observe that guards
in \smartrefs{eq:s:sendSync,eq:s:receiveSync} establish the mandatory
conditions needed for the sender/receiver processes to communicate via
the synchronous channel.  At system level, functions
$\mathoper{wantsynch}$ and $\mathoper{synch}$, given in
Section~\ref{sec:programDen}, make use of these guards to guarantee
that both processes can only progress synchronously.

Finally, \smartref{eq:s:run} models a process call of the \emph{run}
statement for the interpretation $\In$.  All conditional traces built
by this rule start with a conditional transition always enabled that
allocates the needed space for the local variables of the new process;
stores the values of the actual parameters into (the locations of) the
formal parameters; creates the value for the $\predef{pid}$ variable
of the new process; and increments the global variable
$\predef{nr\_pr}$ (the number of active processes).  The second
conditional step is a process spawn that contains the suitable
instance of the denotation of the process definition in $\In$, whose
``working locations'' are those settled in the first transition.
The interleaving behavior is handled in a second phase of the
semantics computation.
\begin{example}[Semantics for assignment statement]
    Given an environment $\rho$ and an interpretation $\In$, the
    semantics for an assignment \texttt{x=y} is
    
    \vspace{-3.2ex}
    {\small
    \begin{align*}
        \SemStmt{\mathtt{x=y}}{\rho}{\In} &= \CondStep{\lambda\sigma .\, \True}{\lambda\sigma
        .\, \SemLEval{\mathtt{x}}{\rho}{\sigma}( \lambda v .
        \SemREval{\mathtt{y}}{\rho}{\sigma} ) } 
        = \CondStep{\lambda\sigma .\, \True} {\lambda\sigma .\,
        \sigma\update{\rho(\mathtt{x})}{\sigma(\rho(\mathtt{y}))} }
    \end{align*}
    }%
    This expression can be read as a conditional transition that is
    always enabled and, as expected, transforms the state $\sigma$ by
    copying the value (stored in the location) of variable
    $\mathtt{y}$ (\ie{} $\sigma(\rho(\mathtt{y}))$) to (the location
    of) variable $\mathtt{x}$ (\ie{} $\rho(\mathtt{x})$).
\end{example}

\subsection{Denotation of process definitions}\label{sec:processDen}

Since we have arbitrary goto instructions in the programs, the flow of
control does not follow predefined patterns like when we have only
structured constructs.  To define a semantics by ``straight''
(one-to-one) induction on syntax one has to resort to the use of some
sort of higher-order constructions (like continuations).  However, the
development of abstract semantics based on this type of concrete
semantics certainly becomes more complicated.

In order to build ``first-order'' denotations we choose to define the
semantics of process definitions $\mathit{proctype}~p( \zseq{x}
) \{ \mathit{body}\}$ by induction on the structure of the Control
Flow Graph (CFG) $\Pi_{p}$ of its body.  Such CFG $\Pi_{p}$ can be
easily constructed from the process syntax, likewise the labeled
transition systems of the operational engine.  More specifically, each
node of a CFG is what we call a \emph{process point}, \ie{} a point
that is referred by the process counters of the operational engine as
well as a
\texttt{goto} statement\footnote{Note that due to the operational semantics of \texttt{goto}
statements, nodes in the CFG might not correspond to states of
the \promela{} process since, in the language, process counters never
point to \texttt{goto} statements.  However, because
of \eqref{eq:s:goto}, the steps in traces generated by the semantics
do correspond to system states.  }.
We follow the typical convention to name process points as the
statement labels (when they are present in the program).  Furthermore,
we interchange each label with the corresponding process point.
In the sequel, we denote by $\labs$ the set of process points of
process $p$ (the nodes of $\Pi_{p}$).

Edges of the CFG are labeled by the basic statement that generates
such edge.
The construction of the CFG is standard, except for the case of
statement \texttt{if/fi}.  For a generic $L$:~$\mlst{if
} \big[ \mlst{::} \mathit{grd}_i; L_i\mlst{:} \mathit{stmts}_{i}
\big]_{i \in \Lambda} \big[ \mlst{::else-> } \mathit{eopt;} \mathit{stmts}_{e} \big] \mlst{
fi};~L'$: $\cdots$, for all branches in $\Lambda$ we have an edge
labeled with $\mathit{grd}_i$ from process point $L$ to process point
$L_i$ and then we have the path (starting at point $L_i$)
corresponding to (the translation of) the statements
$\mathit{stmts}_{i}$ (if the last statement of $\mathit{stmts}_{i}$ is
reached, the path ends in point $L'$).  When the
\mlst{else} branch is present, we add also a path where the first arc is a special one
$\arcElse{\mathit{eopt}}{\bar{l}}$ and then we proceed with (the
translation of) the statements $\mathit{stmts}_{e}$.
In the sequel we denote by $\arcs$ the set of edges $\arc{s}{l'}$ or
$\arcElse{s}{l'}$ of $\Pi_{p}$.
\begin{example}\label{ex:CFGpeterson}
    The CFG of the process {\tt P} defined in \smartref{fig:peterson}
    is the following.
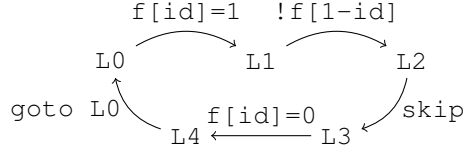
\begin{center}
            \begin{tikzpicture}[scale=1]
                \node (st) at (-1,1) {} ;
                \node (L0) at (0,1) {\texttt{L0}} ;
                \node (L1) at (2,1) {\texttt{L1}} ;
                \node (L2) at (4,1) {\texttt{L2}} ;
                \node (L3) at (3,0) {\texttt{L3}} ;
                \node (L4) at (1,0) {\texttt{L4}} ;
                
                \path[->] 
                (L0) edge[bend left] node[above]{\texttt{f[id]=1}} (L1)     
                (L1) edge[bend left] node[above]{\texttt{!f[1-id]}} (L2) 
                (L2) edge[bend left] node[right]{\texttt{skip}} (L3)
                (L3) edge node[above]{\texttt{f[id]=0}} (L4) 
                (L4) edge[bend left] node[left]{\texttt{goto L0}} (L0) ;
            \end{tikzpicture}
        \end{center}
\end{example}

The denotation of a process definition is built by using the following
immediate consequences operator $\SemTP[]{}{}{}
: \mathit{procs} \times \mathit{Env} \times \Cinterps \times [ \labs
\to \C ] \to [ \labs \to \C ]$.
This operator uses a family of sets of conditional traces indexed by
process points (coded with a function $\eta \in [ \labs \to \C ]$)
that convey the behavior from each process point ``to the end'' and
then builds (backward) the behaviour of all statements that after one
execution step proceed like specified by $\eta$.
{\small
\begin{equation*}
    \SemTP{\rho}{\In}{\eta} \dfn \lambda l .\, 
    \begin{cases}
        \mathoper{addElse}\big( \SemStmt{s'}{\rho}{\In} \cdot \eta(l'') , \Club{_{
        \arc{s}{l'}\in\arcs} \SemStmt{s}{\rho}{\In} \cdot \eta(l') }{} \big) & \text{if }
        \arcElse{s'}{l''} \in \arcs \\
        \Club{_{ \arc{s}{l'}\in\arcs} \SemStmt{s}{\rho}{\In} \cdot
        \eta(l') }{} & \text{otherwise}
    \end{cases}
\end{equation*}
}%
where {\small $\mathoper{addElse}(E,T) \dfn \Club{T}{ \CondStep{ \big( \bigwedge_{\phi \cdot \psi \in
T} \neg \cond{\phi} \big)}{\id} \cdot E }$}.
The $\mathoper{addElse}$ semantic operator is used for \mlst{else}
branches and it adds the negation of each guard of the other branches
in order to make the \mlst{else} branch executable only when all other
branches are not.  Note that conjunction
$\bigwedge_{\phi \cdot \psi \in T}
\neg \cond{\phi}$ is finite since we have finitely many different first conditional states
$\phi$ in any $T$.

Given a process definition $\Pi_{p}$, an environment $\rho$ and an
interpretation $\In$, the function $\SemProc[]{}{}
: \mathit{procs} \times \mathit{Env} \times \Cinterps \to [ \labs \to
\C ]$, written $\SemProc{\rho}{\In}$, builds the conditional traces corresponding to the
executions in isolation of $p$.
\begin{align*}
    \SemProc{\rho}{\In} &\dfn \lambda l .\, \atomics{ \mathcal{F}(l) } && \text{where} &
    \mathcal{F} &\dfn \lfpof{ \lambda\eta .\, \SemTP{\rho}{\In}{\eta}}
\end{align*}

The semantics operator $\atomics{}$ replaces the marked conditional
steps with all possible executions that correspond to the prescribed
semantics of the atomic regions.  It is defined as
$\atomics{\Psi} \dfn \Club{_{\psi\in\Psi} \atomicsAux{\psi} }{}$ where
\begin{itemize}
    \item $\atomicsAux{\psi} \dfn \{ \psi \}$ if $\psi$ does not contain marked conditional steps.
    
    \item Otherwise, without loss of generality, let $\psi = \psi' \cdot \phi_{1} \cdot
    \rho_{1} \cdot \marked{\phi_{2}} \cdot \marked{\rho_{2}} \cdot \psi''$ where $\phi_{1},
    \marked{\phi_{2}}$ are conditional transitions; $\rho_{1}, \marked{\rho_{2}}$ are a
    possibly empty sequences of process spawns; $\psi' \cdot \phi_{1} \cdot \rho_{1}$ has no
    marks and $\marked{\phi_{2}}, \marked{\rho_{2}}$ are marked.  Then
 {\small   \begin{align*}
        \atomicsAux{\psi} \dfn &\Club{ \atomics{ \psi' \cdot ( \CondStep{\lambda \sigma .
        \cond{\phi_1}(\sigma) \wedge \cond{\phi_2}(\trans{\phi_1}(\sigma))}{\lambda \sigma .
        \trans{\phi_2}(\trans{\phi_1}(\sigma))} ) \cdot \rho_{1} \cdot \rho_{2} \cdot \psi'' }
        }{{}} \\
        & \psi' \cdot (\CondStep{\lambda\sigma.  \cond{\phi_{1}}(\sigma) \wedge \neg
        \cond{\phi_{2}}(\trans{\phi_{1}}(\sigma))}{\trans{\phi_{1}}} ) \cdot \rho_{1} \cdot
        \atomics{ \phi_{2} \cdot \rho_{2} \cdot \psi''}
    \end{align*}}%
\end{itemize}
This operator generates traces for the different scenarios in an
atomic region.  It composes the conditions and effects of two atomic
steps $\phi_1$ and $\phi_2$ in a single step.  But the alternative
trace when $\phi_2$ suspends is also generated
($\cond{\phi_{1}}(\sigma) \wedge
\neg \cond{\phi_{2}}(\trans{\phi_{1}}(\sigma))$).  This strategy is applied recursively.

\begin{example}[Process denotations]
    We start by showing the output of the fixpoint computation for
    the \emph{init} process in \smartref{fig:peterson}.
    $\lfpof{ \lambda\eta
    .\, \SemTP[\Pi_\mathit{init}]{\rho}{\In}{\eta}} (\texttt{L5})$ is
{\small
    \begin{align*}
        & \big(\CondStep{\lambda\sigma.  \True}{\lambda \sigma .  \sigma
        (\update{\zseq{l_1}}{\zseq{v_1}} \update{l_{\mathit{1}}}{\mathit{1}}
        \update{\rho(\mathit{\predef{nr\_pr}})}{\sigma(\rho(\mathit{\predef{nr\_pr}})) +1}}
        \big) \cdot{} \notag \RunAtom{\In(\mathtt{P}(0,\_,1))} \cdot{}\\
        & \big(\CondStepAtom{\lambda\sigma.  \True}{\lambda \sigma .  \sigma
        (\update{\zseq{l_2}}{\zseq{v_2}} \update{l_{\mathit{2}}}{\mathit{2}}
        \update{\rho(\mathit{\predef{nr\_pr}})}{\sigma(\rho(\mathit{\predef{nr\_pr}}))   +1}}
        \big)
        \cdot{} \notag \RunAtom{\In(\mathtt{P}(1,\_,2))}
        \cdot{} \Stop
    \end{align*}
    }%
    The atomic block ensures that the two processes are created before
    any of them starts executing its body.  Since it is not possible
    that those two \texttt{run} statements suspend, the $\atomics{}$
    function compacts the two steps and prevents the interleaving with
    the two newly created processes.  In fact
    $\SemProc[\Pi_\mathit{init}]{\rho}{\In} (\texttt{L5})$ is
{\small
    \begin{align}
        & \big(\CondStep{\lambda\sigma.
        \True}{\lambda \sigma .  \sigma (\update{\zseq{l_1}}{\zseq{v_1}}
        \update{l_{\mathit{1}}}{\mathit{1}} \update{\zseq{l_2}}{\zseq{v_2}}
        \update{l_{\mathit{2}}}{\mathit{2}}
        \update{\rho(\mathit{\predef{nr\_pr}})}{\sigma(\rho(\mathit{\predef{nr\_pr}})) +2}}
        \big) \cdot{}\notag \\ &\quad \Run{\In(\mathtt{P}(0,\_,1))} \cdot{}
        \Run{\In(\mathtt{P}(1,\_,2))} \cdot{} \Stop\label{eq:SemProcessInit}
    \end{align}
    }%
The fixpoint computation for the \texttt{P} process type results in a
    single infinite trace in which the four conditional steps actually
    shown below are infinitely repeated:
    {\small
    \begin{align*}
        & \lfpof{ \lambda\eta .\, \SemTP[\Pi_\mathit{P}]{\rho}{\In}{\eta}}(\texttt{L0}) =
        \big\{
        \begin{aligned}[t]
            & \CondStep{\lambda\sigma.  \True}
            {\lambda\sigma.\sigma\update{\sigma(\rho(\mlst{f[id]}))}{1}}
            \cdot {} \CondStepAtom{\lambda\sigma.  \sigma(\rho(\mlst{!f[1-id]}))} {\mathit{id}}
            \cdot {} \\
            & \CondStep{\lambda\sigma.\True}{\id} \cdot{} \CondStep{\lambda\sigma.  \True}
            {\lambda\sigma.\sigma\update{\sigma(\rho(\mlst{f[id]}))}{0}}\cdot\cdots\big\}
        \end{aligned}
    \end{align*}
    }%
    
    The second step in the loop is marked as atomic whereas the third
    one corresponds to the execution of the \texttt{skip} statement
    representing the critical section.  The atomic region in this
    process prevents any interleaving between the first and second
    step if the condition of the second step is satisfied.  However,
    if that condition does not hold, then interleaving is possible.
    This is why the \atomics{} function generates an infinite set of
    infinite traces.  The following set only shows the two possible
    beginnings for traces in
    $\SemProc[\Pi_P]{\rho}{\In}(\texttt{L0})$.  Each trace will
    consist of an arbitrary combination of these fragments.
{\small
    \begin{align}
        & \big\{
        \begin{aligned}[t]
            & \CondStep{\lambda\sigma.  \sigma(\rho(\mlst{!f[1-id]}))}
            {\lambda\sigma.\sigma\update{\sigma(\rho(\mlst{f[id]}))}{1}}
            \cdot {} \\
            & \quad
            \CondStep{\lambda\sigma.\True}{\id} \cdot{}
            \CondStep{\lambda\sigma.  \True}
            {\lambda\sigma.\sigma\update{\sigma(\rho(\mlst{f[id]}))}{0}}\cdot\cdots,\\
            & \CondStep{\lambda\sigma.  \sigma(\rho(\mlst{f[1-id]}))}
            {\lambda\sigma.\sigma\update{\sigma(\rho(\mlst{f[id]}))}{1}}
            \cdot {} 
            \CondStep{\lambda\sigma.  \sigma(\rho(\mlst{!f[1-id]}))}
            {\mathit{id}}
            \cdot {} \\
            &\quad\CondStep{\lambda\sigma.\True}{\id}\cdot{}\CondStep{\lambda\sigma.  \True}
            {\lambda\sigma.\sigma\update{\sigma(\rho(\mlst{f[id]}))}{0}}\cdot\cdots\,,\,%
            \ldots
            \big\}
        \end{aligned}\label{eq:SemProcessP}
    \end{align} 
    }
\end{example}

\subsection{Denotation of a \promela{} Program}\label{sec:programDen}

We are ready to provide the semantic function that computes the traces
associated with a
\promela{} program.  In order to simplify the definitions but without loss of generality, we
assume that programs accepted by this function are normalized.
Essentially, for all processes declared as active we add a suitable
run statement in the init process, encapsulated into an atomic block.
Also, we add an initialization to the proper default value to all
local variable declarations without an initialization.  We move all
global variable declarations to the top and remove \emph{printf}
statements.

In order to define the semantic function for programs, let us start by
defining the denotation for process declarations.
Given a list of process declarations, an environment, and an
interpretation, the function $\SemProcs{}{}{}
: \mathit{Proctypes} \times Env \times \Cinterps \to \Cinterps$
applies the effect of one step of the computation to the input
interpretation.
{\small
\begin{align}
    \label{eq:Dsemantics} & \SemProcs{\mathit{proctype} \: p_{1}( \zseq{x_{1}} ) \{ L_{1}:
    \mathit{body}_{1}\} ;\, \ldots ;\, \mathit{proctype} \: p_{n}( \zseq{x_{n}} ) \{ L_{n}:
    \mathit{body}_{n} \} } {\rho_g}{\In} \dfn \\*&\qquad\notag
    \begin{cases}
        p_{1}( \zseq{l_1}, \zseq{l^v_1}, l^p_1 ) \mapsto \SemProc[\mathit{body}_{1}]{
        \rho_{1}}{\In} (L_{1}) \cdot \mathit{End} & \\
        \vdots & \\
        p_{n}( \zseq{l_n}, \zseq{l^v_n}, l^p_n ) \mapsto \SemProc[\mathit{body}_{n}]{
        \rho_{n}}{\In} (L_{n}) \cdot \mathit{End} & \\
    \end{cases}
    \\*&\notag \text{where }
    \begin{aligned}[t]
        & \rho_{i} \dfn \rho_g \update{\zseq{x_i}}{\zseq{l_i}}
        \update{\zseq{v_i}}{\zseq{l^v_i}} \update{\predef{pid}}{l^p_i} ,\, \qquad
        \zseq{v_i}\text{ are the local variables of } \mathit{body}_{i} ,\\
        & \mathit{End} \dfn \{ \CondStep{\lambda\sigma.  \True}
        {\lambda\sigma .  \sigma \update{\rho(\predef{nr\_pr})}
        {\sigma(\rho(\predef{nr\_pr}))-1} } \cdot \Stop \}
    \end{aligned}
\end{align}
}%
$\mathit{End}$ is the set of traces representing the final step of
each process, \ie{} is responsible for decreasing the counter of
process instances and then stop.

We can now introduce $\SemProg{}{}
: \mathit{Prog} \times \mathit{State} \to \mathbb{T}$ where
$\mathbb{T}$ is the domain of sets of sequences of states in
$\mathit{State}$.  Note that the definition uses two auxiliary
functions $\propagate{}$ and $\interlv{}$ that are formalized later.
Given a normalized program $P\in\mathit{Prog}$ with init process
$\mathit{init}$, global variable declarations $\mathit{gv}$ and a list
of process declarations $\mathit{pd}$, we define
\begin{multline}
    \SemProg{\mathit{gv};\mathit{pd}}{\sigma} \dfn \propagate[\Big]{ \sigma ,\, \interlv[\Big]{
    \big( \Club{ _{k\in\mathbb{N}} ( \lambda \In .  \SemProcs{\mathit{pd}} {\rho_v}{\In} ) ^k
    (\lambda \kappa.\, \Cbot) }{} \big) (\mathit{init}(l_0)) } }
    \label{eq:SemProg}
\end{multline}
where $l_0$ is the location reserved for the \predef{pid} variable of
$\mathit{init}$ and $\rho_v$ is the environment populated with
associations, of both the language predefined variables and the global
variables declared in $\mathit{gv}$, to suitable (different fresh)
locations.

$\SemProg{}{}{}$ first computes the least fixpoint $\mathcal{F}$ of
the function $\SemProcs{}{}{}$; the result is used to compute the
interleaving of spawned processes, and finally the initial state
$\sigma$ is propagated to the conditional traces of
$\mathcal{F}(\mathit{init})$.
$\SemProg{}{}$ is well defined since $\SemProcs{}{}{}$ is continuous.

Given a set of conditional traces, $\interlv{} : \C \rightarrow \C$
replaces the process spawns $\Run{\Psi}$ of each conditional trace
with all possible interleavings of $\Psi$ with the rest of the trace.
To define $\interlv{\Psi}$, first observe that if $\Psi$ is not
$\Cbot$ then it can be partitioned depending on the initial
conditional step of all its conditional traces $\phi_{1},
\ldots \phi_{n}$.  Then
{\small
\begin{align*}
    & \interlv{ \phi_{1} \cdot \Psi_{1} \cup \ldots \cup \phi_{n} \cdot \Psi_{n}} = \Club{
    \Club{_{1\leq i \leq n} \interlvAux[\big]{ \phi_{i}, \Psi_{i} \cup \bigcup_{j\neq i}
    \phi_{j} \cdot \Psi_{j} } }{} }{{}} {} \\
    & \quad \Club{ \set*{ \Club{ \mathoper{synch}(\phi_{j},\phi_{k}) \cdot \Psi }{
    \mathoper{synch}(\phi_{k},\phi_{j}) \cdot \Psi } }{ 
    \begin{aligned}
        & \mathoper{wantsynch}(\phi_{j},\phi_{k}) ,\, \\
        & \Psi = \interlv[\Big]{ \Psi_{j} \cup \Psi_{k} \cup \bigcup_{i\neq j ,\, i\neq k}
        \phi_{i} \cdot \Psi_{i} }
    \end{aligned}
    } }{}
    \\*&\notag \text{where }
    \begin{aligned}[t]
        & \interlvAux{ \Stop , \Psi } = \Psi ,\qquad \interlvAux{ \CondStep{c}{t} , \Psi } =
        \CondStep{c}{t} \cdot \interlv{\Psi} ,\\
        & \interlvAux{ \Run{\Psi'}  , \Psi } = \interlv{\Psi \cup \Psi'} 
    \end{aligned}
\end{align*}
}%
The second part of the definition uses predicate
$\mathoper{wantsynch}(\phi_{s},\phi_{r})$ to detect if conditional
state $\phi_{s}$ wants to perform a synchronous send over a channel,
and $\phi_{r}$ wants to perform a synchronous receive over that
channel.  Moreover $\mathoper{synch}(\phi,\phi')$ is the conditional
step that passes the message content and combines the effects on the
state of $\phi$ and then $\phi'$.  Both actual combinations
$\mathoper{synch}(\phi_{s},\phi_{r})$ and
$\mathoper{synch}(\phi_{r},\phi_{s})$ are possible.
  
Given a set $\Psi$ of conditional traces \emph{not containing process
spawns}, semantic function $\propagate{}
: \mathit{State} \times \C \to \mathbb{T}$ propagates the initial
state $\sigma$ through each conditional trace in $\Psi$, giving as a
result sequences of states.  If $\Psi$ is not $\Cbot$ then it can be
partitioned depending on the initial conditional step of all its
conditional traces.  By construction we can possibly have just one
trace that starts with $\Stop$.  Thus all traces of
$\Psi \setminus \{\Stop\}$ must begin with a finite number of
conditional transitions.  Thus we can formally define $\propagate{}$
as {\small
\begin{align*}
    & \propagate{\sigma ,\, \{\Stop\} \cup \Psi } = \{\xi\} \cup \propagate{\sigma ,\, \Psi }
    \\
    & \propagate{\sigma, \phi_1 \cdot \Psi_1 \cup \dots \cup \phi_k \cdot \Psi_k} = \bigcup
    \set[\big]{ \trans{\phi_i}(\sigma) \cdot \propagate{\trans{\phi_i}(\sigma), \Psi_i } }{
    \cond{\phi_i}(\sigma) = \True}
\end{align*}
}%
At each step, the next conditional transition of each trace is
evaluated, obtaining a new state in the collected sequence.  When a
trace has no conditional transition it is not considered as input for
the following iteration.
Note that the second equation characterizes traces whose conditions
become true after propagation. Since the \interlv{} function generates
non-real traces, we cannot distinguish blocking behaviors from wrong
behaviors and we discard any of them.

Let us illustrate the semantics computation by means of an example.

\begin{example}[Computation of the program semantics]
    Let $P$ the program in \smartref{fig:peterson} and
    $\mathit{body_\mathit{init}}$ the body of the init process.  Let
    $D$ be the process declarations, \ie{}
    $D=\mathtt{init\{}\mathit{body_\mathit{init}}\}\mathtt{; proctype\
    P(bit\ id)\{\mathit{body_p}\}}$ and $\Pi_\mathit{init}$, $\Pi_p$
    their associated CFGs.  Then, the semantics of the whole program
    is
\begin{align*}
        & \SemProg{P}{\sigma_0} \dfn  \propagate[\Big]{ \sigma_0 ,\,
        \interlv[\Big]{ \big( \Club{ _{k\in\mathbb{N}} ( \lambda \In .  \SemProcs{D}
        {\rho_G}{\In} ) ^k (\lambda \kappa.\, \Cbot) }{} \big) (\mathit{init}(l_0)) } } 
    \end{align*}
    where the global environment $\rho_G$ binds \mlst{f[0]} to
    $l_{f0}$, \mlst{f[1]} to $l_{f1}$, \mlst{id} to $l_\mathit{id}$,
    and \predef{nr\_pr} to $l_c$.  Moreover $l_0$ is the location
    reserved for the \predef{pid} variable of $\mathit{init}$.
    Let
    $\rho_{p} \dfn \rho_G\update{\predef{pid}}{l_p}\update{\mathtt{id}}{l_{\id}}$
    and $\rho_{i} \dfn\rho_G\update{\predef{pid}}{l_p}$.
    By \smartref{eq:Dsemantics},
    {\small
    \begin{align*}
        & \SemProcs{D} {\rho_G}{\In} =
        \begin{cases}
            p(\texttt{id},\_,l_p) \mapsto \SemProc[\Pi_p]{\rho_{p}}{\In} (\mathtt{L0}) \cdot
            \mathit{End} & \\
            \mathit{init}( l_p ) \mapsto \SemProc[\Pi_{\mathit{init}}] {\rho_{i}}{\In} \cdot
            \mathit{End} & \\
        \end{cases}
    \end{align*}
    }%
    
    We now illustrate how the unrolling process of the trace generated
    by the init process works.  For conciseness, let $\phi_s$ be the
    starting conditional step {\small
    $\CondStep{\lambda\sigma.  \True}{\lambda \sigma
    .  \sigma \update{\zseq{l_1}}{\zseq{v_1}} \update{l_{\mathit{1}}}{\mathit{1}} \update{\zseq{l_2}}{\zseq{v_2}} \update{l_{\mathit{2}}}{\mathit{2}} \update{\rho(\mathit{\predef{nr\_pr}})}{\sigma(\rho(\mathit{\predef{nr\_pr}}))
    +2}}$ }.  Then,
{\small
    \begin{align*}
        & \interlv[\Big]{\big\{\phi_s\cdot 
        \Run{\In(\mathtt{P}(0,\_,1))}
        \cdot{}
        \Run{\In(\mathtt{P}(1,\_,2))}
        \cdot{} \mathit{End})\big\}} =\\[-1ex]
        &  \big\{\phi_s 
        \cdot{} \interlv[\Big]{\{\Run{\In(\mathtt{P}(0,\_,1))}
        \cdot{}
        \Run{\In(\mathtt{P}(1,\_,2))}
        \cdot{} \mathit{End}\}}\big\}
        = \\[-1ex]
        &  \big\{\phi_s\cdot{} \interlv[\Big]{
        \{\Run{\In(\mathtt{P}(1,\_,2))}
        \cdot{} \mathit{End}\}\cup D_{p1}}\big\}
        = 
        \big\{\phi_s\cdot{} \big\{\interlv[\Big]{ \{\mathit{End}\} \cup D_{p1} \cup D_{p2}\big\}}\cup \Psi \big\}
    \end{align*}
    }%
where $D_{p1}$ and $D_{p2}$ are the sets of traces
    for \mlst{P(0,\_,1)} and \mlst{P(1,\_,2)} resulting
    from \smartref{eq:SemProcessP}, and $\Psi$ represents the rest of
    the unrolling of the traces in $D_{p1}/D_{p2}$. Below we show the
    set $D_{p1}$. $D_{p2}$ is similar to $D_{p1}$, substituting
    every \mlst{f[0]} by \mlst{f[1]} and every \mlst{f[1]}
    by \mlst{f[0]}.
    {\small
    \begin{align*}
        &  D_{p1} =  \big\{
        \begin{aligned}[t]
            & \CondStep{\lambda\sigma.  \sigma(\rho(!f[1]))}
            {\lambda\sigma.\sigma\update{\sigma(\rho(\mlst{f[0]}))}{1}}
            \cdot {} \CondStep{\lambda\sigma.\True}{\id} \cdot{} 
            \CondStep{\lambda\sigma.  \True}
            {\lambda\sigma.\sigma\update{\sigma(\rho(\mlst{f[0]}))}{0}}\cdot\cdots,\\
            & \CondStep{\lambda\sigma.  \sigma(\rho(f[1]))}
            {\lambda\sigma.\sigma\update{\sigma(\rho(\mlst{f[0]}))}{1}}
            \cdot {} 
            \CondStep{\lambda\sigma.  \sigma(\rho(!f[1]))}
            {\mathit{id}}
            \cdot {} \\
            &\quad \CondStep{\lambda\sigma.\True}{\id} \cdot{}
            \CondStep{\lambda\sigma.  \True}
            {\lambda\sigma.\sigma\update{\sigma(\rho(\mlst{f[0]}))}{0}}\cdot\cdots\,,\,
            \ldots
            \big\}
        \end{aligned}
    \end{align*}
    \vspace{-1ex}
    }%
    
    The \interlv{} function takes each trace in these sets and
    computes all possible interleavings between each pair of traces.
    We show the beginning of two computed traces, one in which the
    first process executes once the loop followed by the execution of
    the loop by the second process, and another in which after the
    first two steps of the first process, the second one executes its
    first condition and then assumes the second step is not
    executable.
    {\small
    \begin{align}\label{eq:interleaving}
        &\phi_s
        \cdot{}\\*&\, \notag \{
        \begin{aligned}[t]
            & \CondStep{\lambda\sigma.  \sigma(\rho(!f[1]))}
            {\lambda\sigma.\sigma\update{\sigma(\rho(\mlst{f[0]}))}{1}}
            \cdot {} \CondStep{\lambda\sigma.\True}{\id} \cdot{} 
            \CondStep{\lambda\sigma.  \True}
            {\lambda\sigma.\sigma\update{\sigma(\rho(\mlst{f[0]}))}{0}}\cdot{}\\\notag
            &\quad  \CondStep{\lambda\sigma.  \sigma(\rho(!f[0]))}
            {\lambda\sigma.\sigma\update{\sigma(\rho(\mlst{f[1]}))}{1}}
            \cdot {} \CondStep{\lambda\sigma.\True}{\id} \cdot{}
            \CondStep{\lambda\sigma.  \True}
            {\lambda\sigma.\sigma\update{\sigma(\rho(\mlst{f[1]}))}{0}}\cdot\cdots,\\\notag
            & \CondStep{\lambda\sigma.  \sigma(\rho(!f[1]))}{\lambda\sigma.\sigma\update{\sigma(\rho(\mlst{f[0]}))}{1}} \cdot{}
            \CondStep{\lambda\sigma.\True}{\id} \cdot{}
            \CondStep{\lambda\sigma.  \sigma(\rho(f[0]))}{\lambda\sigma.\sigma\update{\sigma(\rho(\mlst{f[1]}))}{1}} \cdot{}\\\notag
            & \quad 
            \CondStep{\lambda\sigma.  \True}
            {\lambda\sigma.\sigma\update{\sigma(\rho(\mlst{f[0]}))}{0}}\cdot{}
            \CondStep{\lambda\sigma.  \sigma(\rho(f[1]))}{\lambda\sigma.\sigma\update{\sigma(\rho(\mlst{f[0]}))}{1}} \cdot{}
            \CondStep{\lambda\sigma.  \sigma(\rho(!f[0]))}{\id} \cdot\cdots,\\
            & \cdots\}
        \end{aligned}
    \end{align}
    }
    
    After the interleaving process, the propagate operator accumulates
    the effects of the executed steps.  For conciseness, let
    $\sigma_{(a,b)}$ be the state $\langle \{l_0\mapsto a,l_1\mapsto
    b\}, \{f[0]\mapsto l_0,f[1]\mapsto l_1\}\rangle$.  The semantics
    for our program example includes all traces in which each process
    runs the complete loop before passing the control (or not) to the
    other process:
    {\small
    \begin{align*}
        & \propagate[\Big]{ \sigma_{(0,0)} ,\,
        \interlv[\Big]{\mathit{init}(l_0)}} = 
        \{\begin{aligned}[t]
        & \sigma_{(0,0)} \cdot{} \sigma_{(1,0)} \cdot {}  \sigma_{(1,0)} \cdot {} 
        \sigma_{(0,0)} \cdot {} \sigma_{(0,1)} \cdot {}\sigma_{(0,1)} \cdot {} \sigma_{(0,0)} \cdots,\\
        & \sigma_{(0,0)} \cdot{} \sigma_{(0,1)} \cdot {}  \sigma_{(0,1)} \cdot {} 
        \sigma_{(0,0)} \cdot {} \sigma_{(1,0)} \cdot {}\sigma_{(1,0)} \cdot {} \sigma_{(0,0)} \cdots,\\
        & \sigma_{(0,0)} \cdot{} \sigma_{(1,0)} \cdot {}  \sigma_{(1,0)} \cdot {} 
        \sigma_{(0,0)} \cdot {} \sigma_{(1,0)} \cdot {}\sigma_{(1,0)} \cdot {} \sigma_{(0,0)} \cdots,
        \ldots\}
    \end{aligned}
\end{align*}
}%

Note that when propagating the second trace
in \smartref{eq:interleaving}, we have that we reach a point in which
the condition of the following step is false, thus the behavior trace
is discarded.  This particular trace, included in the set of traces
generated by the $\interlv{}$ function, corresponds with a deadlocking
trace in \promela{}, namely, $\sigma_{(0,0)} \cdot
\sigma_{(1,0)} \cdot{} \sigma_{(1,0)} \cdot{} \sigma_{(1,1)} \cdot \sigma_{(0,1)} \cdot
\sigma_{(1,1)}$.  The (propagated) state $\sigma_{(1,1)}$ falsifies the conditional trace
$\CondStep{\lambda\sigma.  \sigma(\rho(!f[0]))}{\id}\cdot{}\cdots{}$.
\end{example}

\section{Conclusion and future work}\label{sec:Conclusion}

This work presents a first proposal of a inductive bottom-up fixpoint
semantics for \promela{}, which is a widely used modeling language for
concurrent, reactive systems.  It does not cover the full language,
but the considered fragment is very significant.  In particular, it
handles arbitrary \texttt{goto}s, which are extensively used by
the \promela{} community and both synchronous and asynchronous
communications.
The presented semantics enjoys the good properties to be the basis for
the implementation of verification and analysis techniques.  First, it
is goal-independent, which allows us to collect all possible behaviors
from most general calls, avoiding the need of computing the semantics
individually for each possible goal.  Second, it is bottom-up which
improves both convergence and precision in the abstract-interpretation
setting for analysis and verification.  This is because
the \emph{join} operator of the abstract domain is used less times
than in a top-down approach.

As noted, the denotational semantics in this work does not capture
deadlocking traces.  As a future work we plan to overcome this
limitation by defining a new $\interlv{}$ semantic operator that
propagates the conditions of previous conditional states within the
conditions of successive ones, in order to discard impossible
sequentializations and retain only the finite traces that stop either
on successful termination or deadlock.

\end{document}